\documentstyle[aps,prl,psfig,bm]{revtex}	 
\begin{document}

\title{
Sensitivity of the Mott Transition to Non-cubic Splitting of the 
Orbital Degeneracy: Application to NH$_3$K$_3$C$_{60}$
}

\author{Nicola Manini$^{1,2,3,4}$\thanks{E-mail: nicola.manini@mi.infm.it},
        Giuseppe E. Santoro$^{3,4}$\thanks{E-mail: santoro@sissa.it},
\\
	Andrea Dal Corso$^{3,4}$\thanks{E-mail: dalcorso@sissa.it},
	and Erio Tosatti$^{3,4,5}$\thanks{E-mail: tosatti@sissa.it}
\\ \it 
$^1$ Dip.\ Fisica, Universit\`a di Milano, Via Celoria 16 - 20133
Milano, Italy
\\ \it 
$^2$ INFM, Unit\`a di Milano, Milano, Italy
\\ \it 
$^3$ INFM, Unit\`a Trieste SISSA, Italy
\\ \it 
$^4$ International School for Advanced Studies (SISSA),
\\ \it
Via Beirut 4, I-34013 Trieste, Italy
\\ \it 
$^5$ International Centre for Theoretical Physics (ICTP), 
\\ \it
P.O. Box 586, I-34014 Trieste, Italy
}

\date{21-12-2001}
\maketitle

\begin{abstract}
Within dynamical mean-field theory, we study the metal-insulator
transition of a twofold orbitally degenerate Hubbard model as a function
of a splitting $\Delta$ of the degeneracy. The phase diagram in 
the $U-\Delta$ plane exhibits two-band and one-band metals,
as well as the Mott insulator. The correlated two-band metal 
is  easily driven to the insulator state by a strikingly weak 
splitting $\Delta\ll W$ of the order of the Kondo-peak width $zW$,
where $z\ll 1$ is the metal quasiparticle weight. The possible
relevance of this result to the insulator-metal transition in 
the orthorhombic expanded fulleride NH$_3$K$_3$C$_{60}$ is discussed.
\end{abstract}

\section{Introduction}

Strong electron correlations in multi-band, orbitally
degenerate systems represent an important current theoretical challenge.
A lively experimental playground for that is provided by electron- and
hole-doped fullerene systems, which exhibit a variety of behavior, including
unconventional metals like cubic CsC$_{60}$\cite{Brouet99}, superconductors
of the A$_3$C$_{60}$ family (A= K,Rb,Cs)\cite{Ramirez,Gunnarsson97},
superconducting FET
devices\cite{schon00,Batlogg00}, and insulators, presumably of the
Mott-Jahn-Teller type\cite{Fabrizio97,Capone00}. 
Among them we can place A$_4$C$_{60}$ \cite{Benning},
Na$_2$C$_{60}$\cite{Brouet01} and the class of compounds
NH$_3$K$_3$C$_{60}$, NH$_3$K$_2$RbC$_{60}$, NH$_3$KRb$_2$C$_{60}$, as
well as NH$_3$Rb$_3$C$_{60}$\cite{Rosseinsky93,Takenobu00,Obu00}.
In these ammoniated compounds, insertion of the electronically inert NH$_3$
molecules acts to expand the C$_{60}$ lattice, turning the cubic, metallic and
superconducting state of K$_3$C$_{60}$ into an orthorhombic narrow-gap
antiferromagnetic insulator \cite{Rosseinsky93,Takenobu00,Tou1}. 
%
In this transition, the increase in volume per C$_{60}$ molecule 
relative to K$_3$C$_{60}$, with its probable slight decrease of electron
effective bandwidth $W$, is believed to play an important role, as 
confirmed by the pressure-induced reversion to a
fully metallic and superconducting -- while still orthorhombic -- phase
\cite{Zhou95,Margadonna01}.

Nonetheless, a possible role of the crystal-structure anisotropy 
in this Mott transition -- a transition generally believed to occur 
more readily for reduced orbital degeneracy\cite{Gunnarsson96,Koch99}, 
cannot be excluded.
It has even been proposed that precisely the splitting of 
degeneracy\cite{splitting:note} induced by the orthorhombic distortion 
could be the crucial ingredient driving the transition from the 
metal and superconductor to the insulator \cite{Takenobu00}.
Earlier work showed the Mott transition in $d$-orbitally degenerate lattice
models to take place at larger values of $U/W$ for larger degeneracy, 
roughly proportionally to $\sqrt{d}$
\cite{Gunnarsson96,Koch99}.
A strong enough ``crystal-field'' splitting of the threefold degenerate
$t_{1u}$ molecular orbital of C$_{60}$ caused by orthorhombic anisotropy 
could remove one or several bands away from the Fermi level, 
effectively reducing the orbital degeneracy $d$, thus shifting the Mott
transition to a smaller critical value $U_c/W$.
Exploring this concept, the basic question is  how strong this splitting 
must be to promote such an effective reduction of degeneracy.  
In a non-interacting system, that reduction would clearly require a splitting 
magnitude similar to or larger than the full electron bandwidth $W$.
In a strongly-interacting system, it is important to understand 
whether influencing the metal-insulator transition will 
again require anisotropic splittings as large as the bandwidth, 
or else if some smaller energy scale will emerge in its place.

In this paper we exploit dynamical mean field theory (DMFT)\cite{Georges96} to 
answer this question, studying the effects of a band splitting 
on the Mott transition of an orbitally degenerate, strongly-correlated 
metal.
For that purpose we adopt the simplest idealization, namely
a $d=2$ orbitally degenerate Hubbard model with bandwidth $W$
and on-site repulsion $U$, and mimic the noncubic anisotropy by a 
diagonal splitting $\Delta$ between the two bands. 
In this model, we study the $T=0$ phase diagram in the $(U,\Delta)$ plane,
and examine in particular how the Mott transition is influenced by
the splitting.
%
As the aim of the calculation is understanding a simple, and possibly
fairly general physical mechanism, we do not attempt a realistic calculation 
of the kind that was accomplished recently for Ni and Fe \cite{Lichtenstein01}.

The main result found is that the effect is remarkably strong at large $U$, 
where the anisotropy-induced transition between the correlated 
multi-band metal and the Mott insulator takes place very readily, 
in fact for $\Delta \ll W$. 
We interpret that as follows. The metallic band, originally of
width $W$, is renormalized in the strongly correlated metal into 
a narrower ``Kondo-like'' quasiparticle peak of width $zW$, 
with $1\gg z \to 0$ as $U\to U_c$.
Close enough to the isotropic and orbitally degenerate Mott transition 
at $U\lesssim U_c$, the correlated two-band metal can be driven 
effectively one band by a splitting as small as the Kondo
energy scale $\Delta \propto zW$.  
The one-band metal, in turn, has a lower $U_c$, so that the resulting state 
is an insulator.

With this model result in hand, we move on to consider NH$_3$K$_3$C$_{60}$, 
where we carry out a DFT-LDA (density functional theory in the local density
approximation) electronic-structure calculation to extract 
an order of magnitude for the effective orthorhombic splitting parameter 
$\Delta/W$, found to be of order $0.25$. 
%
Despite the obvious differences between the two-band Hubbard model and the
true three band system, we suggest that the orthorhombic
splitting will similarly drive the initially cubic fulleride with a small
quasiparticle residue $z$ from metal to insulator.
%
In turn, in the light of our simple-model phase diagram,
the return to metallicity of the ammoniated orthorhombic fulleride
under pressure leads to an
interesting question regarding the effective number of bands in that metal.
It is finally concluded that anisotropy is likely a relevant
element in promoting the Mott state in this material.

This paper is organized as follows: Sec.~\ref{model:sec} introduces the
model and deals with its simple limiting cases.
The calculation and results within the DMFT are described is
Sec.~\ref{dmft:sec}. 
The $t_{1u}$ band structure of NH$_3$K$_3$C$_{60}$ and its splitting 
due to anisotropy is described in Sec.~\ref{bands:sec}.
The discussion and conclusions are finally presented in
Sec.~\ref{discussion:sec}.

\section{A simplified two-band Hubbard model} \label{model:sec}

We work with the simplest two-band Hubbard model, assuming 
purely diagonal hoppings between orbitals on different sites, the
anisotropic symmetry lowering all embodied in a diagonal 
on-site splitting term.
We write the Hamiltonian as:
\begin{equation} \label{hamiltonian:eq}
H = - t \sum_{\langle i,j\rangle,\sigma} \sum_{\alpha=1,2}  
\left(c^\dagger_{i\alpha\sigma} c_{j\alpha\sigma} + H.c.\ \right)
+ 
\sum_{i} \sum_{\alpha=1,2} \epsilon_\alpha ~ n_{i\alpha}
+
\frac U2 \sum_{i} n_{i}(n_i-1) \;,
\end{equation}
where $c^\dagger_{i\alpha\sigma}$ creates a spin-$\sigma$ electron in orbital
$\alpha$ at site $i$,
$n_{i\alpha}=\sum_{\sigma} c^\dagger_{i\alpha\sigma} c_{i\alpha\sigma}$, 
$n_{i}=\sum_{\alpha} n_{i\alpha}$ is the total number of electrons at
site $i$, and $\langle i,j \rangle$ denotes nearest-neighbor sites.
%
The first term is a standard tight-binding hopping,
simplified to ignore non-diagonal hopping, and also merohedral disorder,
both of them present in the real systems. 
The second term introduces the anisotropic splitting $\Delta$
as a shift of the on-site orbital energy $\epsilon_2=-\epsilon_1=\Delta/2$.
The last term is the Hubbard on-site interaction, which we take in
its simplest form, omitting for example any additional intra-site Hund's rule 
interaction terms.  
We also ignore the on-site electron-vibron couplings and the ensuing
Jahn-Teller effects, even if quite important for other
aspects\cite{Gunnarsson97,Capone01,MTA}.

In the $U=0$ limit, the splitting $\Delta>0$ simply operates a rigid
shift of band 2 upwards with respect to band 1, promoting electron transfer 
from the upper to the lower band.
Above a critical value $\Delta=\Delta_c$, and for a total 
electron filling $n<2$, the upper band is emptied up. 
For example, with two symmetric bands of width $W$ and $n =1$ electrons 
per site, the upper band is emptied above $\Delta_c/W=0.5$.
At zero temperature (where we shall restrict ourselves in this paper) 
and zero interaction, $U=0$, the transition between the 
``two-band metal'' and the ``one-band metal'' is continuous.
Because the topology of the Fermi surface changes,
this transition is accompanied by a weak singularity of the total
energy first described by Lifshitz\cite{Lifshitz60}.
When the electron-electron interaction $U$ is turned on, one expects 
the emptying of the upper band to take place at smaller values 
of $\Delta_c$, owing to the effective band narrowing. Perturbatively 
in $U$ one can show that 
\begin{equation}
\Delta_c(U) = \Delta_c(0) - \gamma\; U + O(U^2) \;,
\label{smallU}
\end{equation}
where the value of the coefficient $\gamma>0$ depends on details of the
bands.  In addition, electron-electron interactions might modify the nature
of the metal-metal transition singularity relative to the non-interacting
case\cite{Katsnelson00}, a point which we will not further address here.

We restrict to a filling of one electron per site, $n=1$,
discarding the trivial case $n=2$ where the large-$\Delta$ state is a
band insulator, as well as noninteger $n$, where insulators are not
possible.
%
The $\Delta=0$ and $\Delta\to\infty$ limits reduce then, respectively, to the
(quarter-filled) two-band and the (half-filled) single-band Hubbard models,
both possessing a metal-insulator transition as a function of $U$
\cite{Rozenberg97,Rozenberg94,Caffarel94,Bulla99}.
In particular, we are not interested in the weak-coupling antiferromagnetic
instability, of the ensuing large-$\Delta$ half-filled band, associated to
nesting in specific hopping structures: we assume a genuine Mott transition
for the half-filled single-band model to occur at a finite $U_c^{\rm
1-band}>0$.

The limit of strong interaction $U\gg W$ is insulating for any value of
$\Delta$. 
This limit can be studied by mapping the model
(\ref{hamiltonian:eq}) onto a spin and orbital exchange Hamiltonian
which reads\cite{Arovas95}:
\begin{equation}
H_{\rm exch} = J \sum_{<i,j>} \left(
{\bf S}_i\cdot {\bf S}_j +
{\bf T}_i\cdot {\bf T}_j +
4\; {\bf S}_i\cdot {\bf S}_j \ \ {\bf T}_i\cdot {\bf T}_j 
\right)
-\Delta \sum_i T^z_i \;,
\label{Hexch}
\end{equation}
where the pseudospin-1/2 operators ${\bf T}_j=1/2 \sum_{\alpha\beta\nu}
c^\dagger_{j\alpha\nu} {\bm \sigma}_{\alpha\beta} c_{j\beta\nu}$ represent
the orbital degrees of freedom, and $J=2t^2/U$.
For $\Delta=0$ this model has been studied both in one\cite{Sutherland75}
and two dimensions \cite{Zhang98}, with suggestions that interesting
spin-liquid physics could be realized.
For our purposes, it suffices to note that the model has no ferro-orbital
instability, and has therefore a finite $q=0$ orbital susceptibility.
As a consequence it takes a nonzero value of $\Delta$ to 
fully orbitally polarize the ground state.
Due to the absence of cross-band terms in the kinetic energy, complete
orbital polarization occurs at a finite $\Delta_c\propto J=2t^2/U$.
For $\Delta\ge \Delta_c$ that ground state is represented 
by a one band Mott insulator plus a totally empty split-off band.

Notice, finally, that for the specific case we are interested ($n=1$
electron in a two-fold degenerate band) no orbital ordering is
present at weak coupling; the possibility of antiferro-orbital ordering
within the Mott insulating phase, on the other hand, depends crucially on
the details of the hoppings and lattice coordination, and it is, in any case,
beyond the scope of any single-site DMFT study.

\section{Phase diagram of the simple model} \label{dmft:sec}

\begin{figure}[t]
\centerline{\psfig{figure=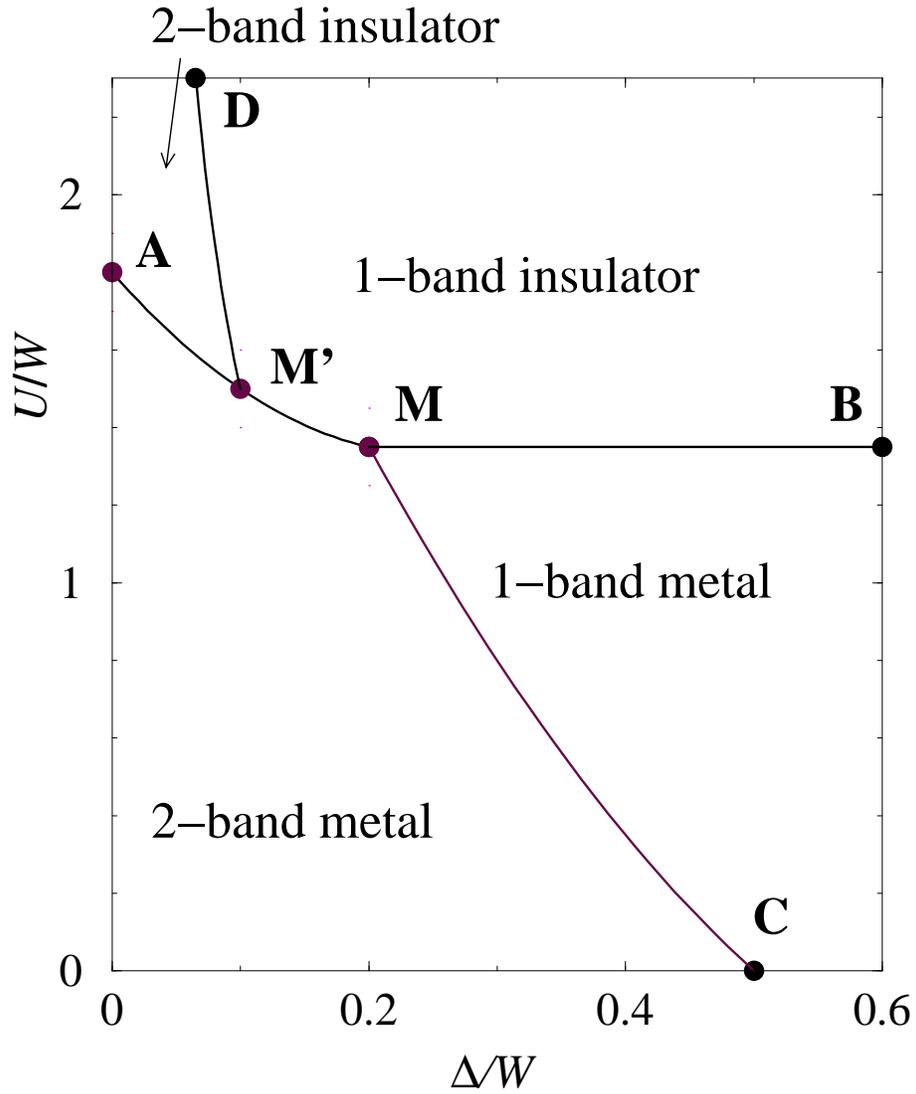,width=12cm}}
\vspace{0.1cm}
\protect\caption{
Qualitative zero-temperature phase diagram for the two-band Hub\-bard model
at quarter filling (one electron per site) in the $U$-$\Delta$ plane, 
where $\Delta$ is the anisotropy splitting of the two orbitals.
The various phases and lines are described in Sect.~\ref{dmft:sec}. 
The multicritical points M and M' are not necessarily distinct.
\label{phased_th:fig}
}
\end{figure}

We have now all the elements to illustrate the general layout, sketched in
Fig.~\ref{phased_th:fig}, of the zero-temperature phase diagram of model
in Eq.\ (\ref{hamiltonian:eq}) as a function of $(U,\Delta)$ for $n=1$ electron per
site.
The metal-insulator transition for $n$=1, $d=2$ and $\Delta=0$, previously
studied in DMFT by quantum Monte Carlo \cite{Rozenberg97}, was found at
$U_c^{\rm 2-band}\approx 1.5\,W$ at the relatively low temperature
$T=W/32$.
The same transition with $n=1$ in the one-band model 
($d=1$, or $d=2$ at $\Delta=\infty$) has been similarly calculated at 
$U_c^{\rm 1-band} \approx 1.3\,W$ also at $T=W/32$\cite{Rozenberg94}.
Finite temperature of course must affect somewhat these numerical values, 
but a common feature with our zero-temperature DMFT calculation 
described below, is that $U_c^{\rm 2-band}>U_c^{\rm 1-band}$.
The AB line in Fig.~\ref{phased_th:fig}, representing $U_c$ as a function
of $\Delta$ and separating metals from Mott insulators, indicates this
trend.
The DM' line separates the fully orbitally polarized Mott insulator 
(at the right) from the two-band insulator, roughly as $U_c\sim\Delta^{-1}$
for small $\Delta$.
Similarly, the CM line separates the fully orbitally polarized 
metal from the two-band metal: it starts from point C with a linear slope
$-\gamma^{-1}$, according to Eq.~(\ref{smallU}).
In the region of full orbital polarization the value of $\Delta$ is
irrelevant, and this is the reason why the Mott transition 
line MB is horizontal.
As $\Delta$ increases, for $U<U_c^{\rm 1-band}$ the upper-band emptying 
transition takes the two-band metal across the CM line over to a one-band 
metal, while for $U>U_c^{\rm 1-band}$ it leads across the AM line to a
Mott insulating state.
The AM line starts off linearly at small $\Delta$.
Basically, this occurs since in close proximity of the Mott transition 
the relevant energy scale, to be matched by the orbital splitting $\Delta$, 
is no longer the bare bandwidth, but rather the width of the Kondo-like 
quasiparticle peak of the correlated metal. 
That in turn is of the order of $zW$, which is smaller than $W$ and 
proportional to the quasiparticle residue $z$, linearly vanishing with 
$(U_c^{\rm 2-band}-U)$ at the Mott transition\cite{Georges96}.
As the DMFT calculations will show, the effective emptying transition
occurs when $\Delta$ increases to reach $\Delta_c(U)\propto zW\propto
(U_c^{\rm 2-band}-U)$.

\begin{figure}[t]
\centerline{\psfig{figure=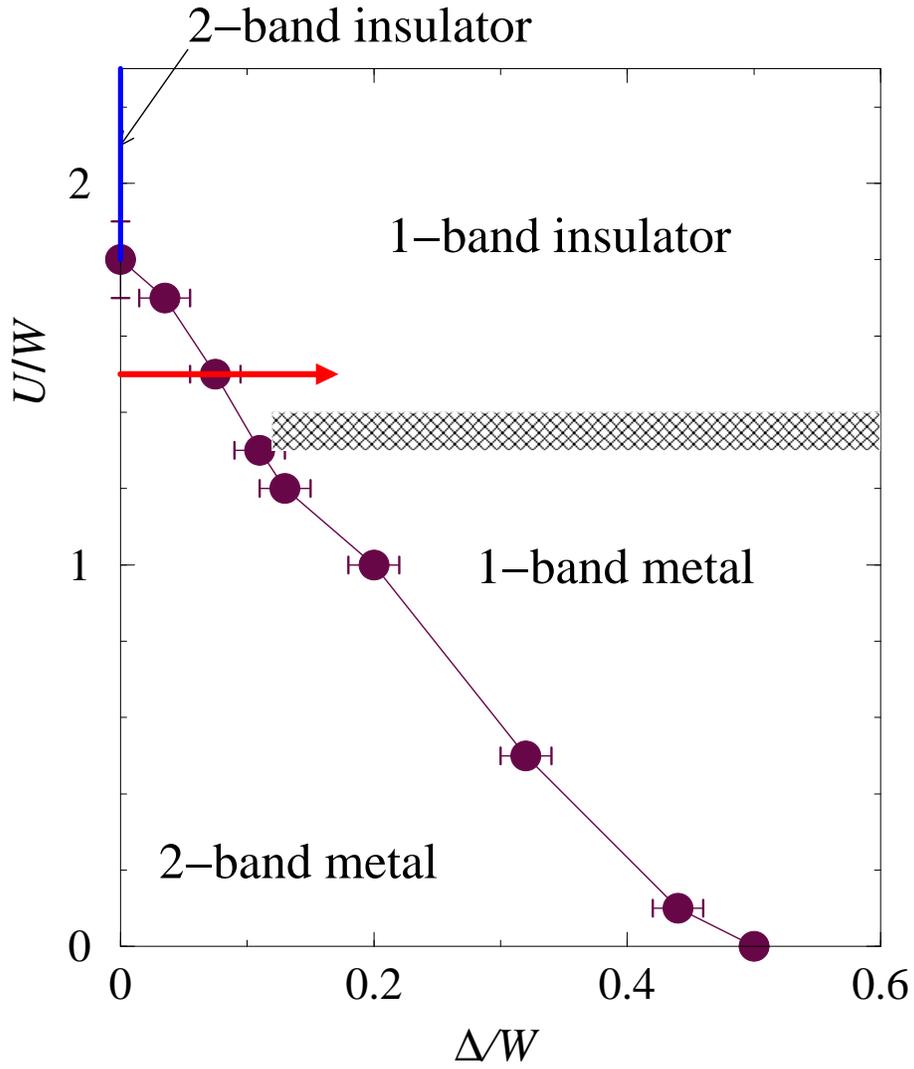,width=12cm}}
\vspace{0.1cm}
\caption{
DMFT zero-temperature phase diagram for the two-band Hub\-bard model
at quarter filling (one electron per site), obtained by the exact 
diagonalization method in the paramagnetic sector. 
The discretization of the conduction band uses here $n_b=4$ bath states per 
orbital degree of freedom.
\label{phase_diagram:fig}
}
\end{figure}

The purpose of the DMFT calculations is to investigate this phase diagram
in greater quantitative detail, in particular in the region of intermediate
$U\sim W$ and $\Delta\lesssim W$.
We have implemented DMFT in the exact diagonalization flavor
\cite{Georges96,Caffarel94} for the degenerate Hubbard model
on the Bethe lattice, restricted
to paramagnetic states \cite{DMFTdetails:note}.
%
The advantage of the exact diagonalization method over the Monte Carlo approach
in the solution of the impurity problem rests primarily in its being less
computationally intensive -- useful when studying the phase
diagram of a model as a function of more than one parameter --
and in its ability to address zero-temperature properties 
\cite{Georges96,DMFTMCdetails:note}.
The results of the DMFT calculations are summarized in 
Fig.~\ref{phase_diagram:fig}.

The values we obtain for $U_c^{\rm 1-band}\simeq 1.35\,W$ and $U_c^{\rm
2-band}\simeq 1.8\,W$, associated to the disappearance of the metallic
state for one and two bands respectively, have been calculated with the
same number of discretized bath levels per band $n_b=4$.
The associated error bar is estimated by repeating the calculation with
$n_b=5$.
These values are in substantial agreement with corresponding values
obtained by other methods\cite{Rozenberg97,Rozenberg94,Caffarel94,Bulla99}.
The other points in the phase diagram are obtained by following the
stability of the two-band metal for a given value of $U$ and increasing
$\Delta$, marking the emptying transition to the one-band metal or to the
insulator.

A deficiency of this single-site DMFT calculation --
which is restricted as usual to 
paramagnetic states only -- is the absence of a two-band insulating state 
for $\Delta>0$. 
In fact, the suppression of the antiferro orbital fluctuations embodied in
the exchange model (\ref{Hexch}) produces a fictitious infinite uniform 
orbital susceptibility which leads to full orbital polarization as soon 
as $\Delta$ is turned on \cite{Georges96}. 
Despite this limitation, the results of the present calculations are
suggestive, revealing the announced sharp reduction of the metal-insulator
$U_c(\Delta)$ for small but finite $\Delta$.
Indeed, our calculations give a $\Delta_c(U)$ which is roughly proportional
to the quasiparticle residue $z(U)$ of the undistorted ($\Delta=0$)
correlated metal at $U<U_c^{\rm 2-band}$,
\begin{equation}
\frac{\Delta_c(U)}{W} \propto \beta z(U) \;, 
\end{equation}
with a proportionality constant $\beta\simeq 0.3$. 
Since $z(U)$ vanishes as $U\to U_c^{\rm 2-band}$, most likely linearly in
$(U_c^{\rm 2-band}-U)$ \cite{Georges96}, even a very small $\Delta/W$ can
be enough to cause a metal-insulator transition in the strongly correlated
metal.
For example, following the bold arrow at $U=1.5\,W$ in 
Fig.~\ref{phase_diagram:fig}, a $\Delta$ value as small as $0.08\, W$ is 
sufficient to cross the transition line from the metal to insulator.

\begin{figure}[t]
\centerline{\psfig{figure=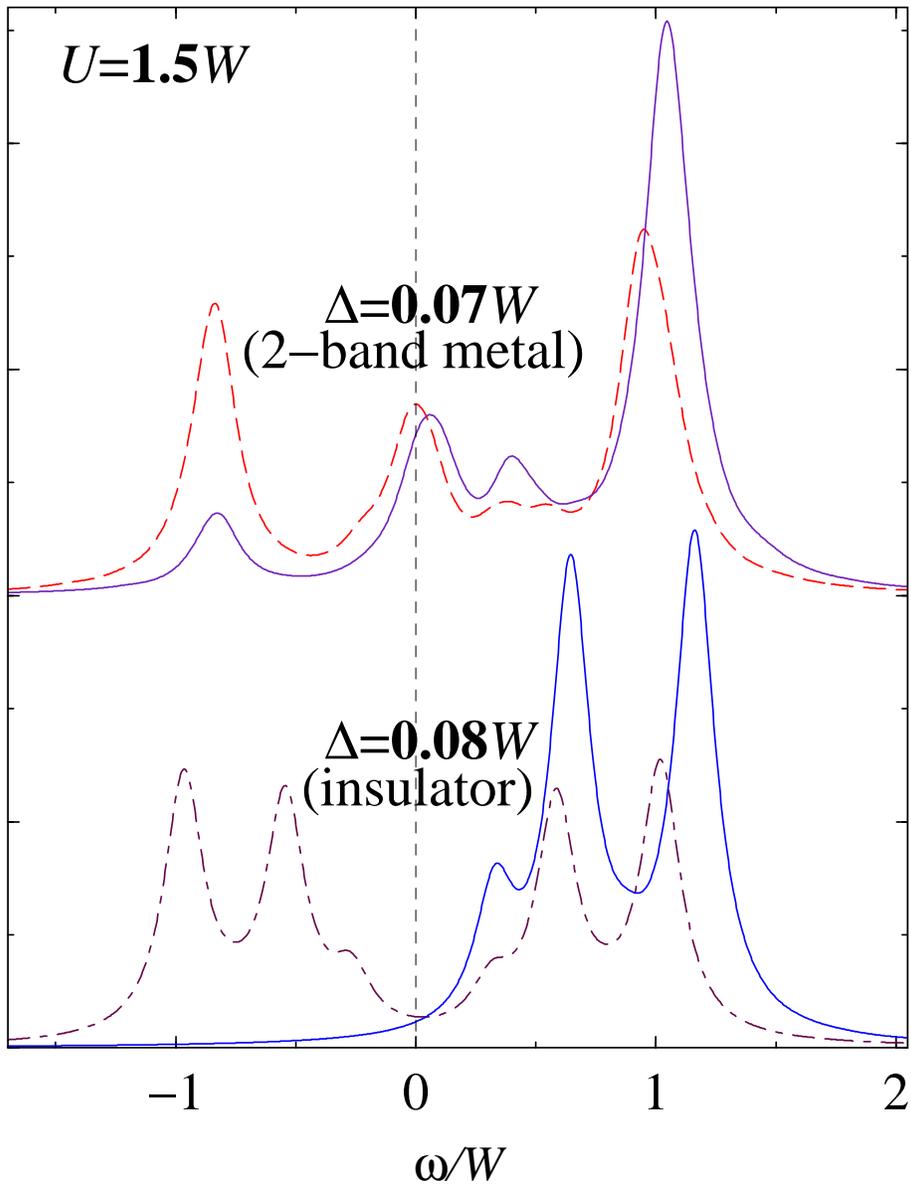,width=12cm}}
\vspace{0.1cm}
\caption{
Spectral density $A_{\alpha}(\omega)=-\pi^{-1} {\rm Im}\,G_{\alpha}(\omega)$ 
at $U/W=1.5$ across the Mott-Hub\-bard transition
for increasing anisotropy splitting $\Delta$.
$\omega$ is referred to the Fermi energy.
Solid lines refer to the minority orbital $\alpha=2$. 
The multi-peak structures of the high-energy side bands are artifacts of
the finite discretization of the conduction band.
\label{spec_dens:fig}
}
\end{figure}

Further illustrating that, in Fig.~\ref{spec_dens:fig} we show the behavior
of the spectral density
$$ A_\alpha(\omega)=-\pi^{-1}\;{\rm Im}\, G_\alpha(\omega)  \;, $$
$G_\alpha(\omega)$ being the one-particle Green's function of band $\alpha$,
on both sides of the metal-insulator transition.
The asymmetry of the upper-band spectral density $A_2(\omega)$ (solid lines) 
is very pronounced, as this band is nearly ($\Delta/W=0.07$) or
completely ($\Delta/W=0.08$) empty.
As soon as the Kondo-like peaks of the two bands differ enough in energy 
to induce the emptying of band 2, the lower-band spectral density 
$A_1(\omega)$ takes the symmetric shape, characteristic of the half-filled 
one-band Hubbard model.
Here the Kondo peak disappears completely, as this value of $U>U_c^{\rm
1-band}$ puts the Hubbard model of band 1 well inside the insulating
regime.

\section{Anisotropic band splitting in orthorhombic NH$_3$K$_3$C$_{60}$} 
\label{bands:sec}

The above model calculations show that a small splitting 
$\Delta \propto  zW$ of the orbitally degenerate band can drive
the metal-insulator transition. 
We now wish to explore the implications that this result -- if assumed to
be more general than the simple model where it was derived -- can have on
the metal-insulator transition which takes place between isoelectronic
K$_3$C$_{60}$ and NH$_3$K$_3$C$_{60}$ (the former cubic and the latter
orthorhombic), and on the insulator-metal transition of NH$_3$K$_3$C$_{60}$
itself under pressure.

\begin{figure}[t]
\centerline{\psfig{figure=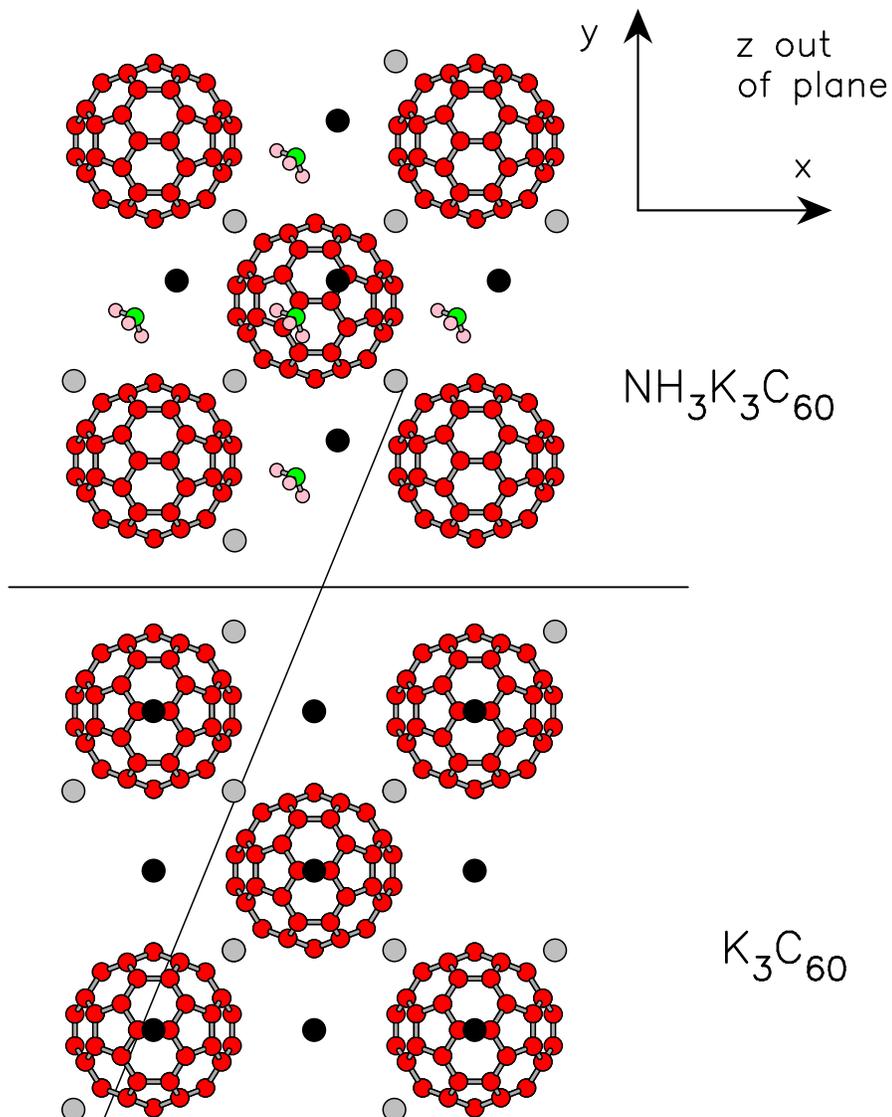,width=12.5cm}}
\vspace{0.1cm}
\caption{Simplified geometry of NH$_3$K$_3$C$_{60}$ (top) and
K$_3$C$_{60}$ (bottom) used in the LDA calculation. Octahedral K atoms
are indicated in black, while tetrahedral ones are in gray. 
All visible atoms of the central unit cell are shown together with some 
atoms in the neighbor unit cells. 
In NH$_3$K$_3$C$_{60}$ the $c$-axis is in the $z$ direction.
\label{geometry_NH3:fig}
}
\end{figure}

To obtain a quantitative estimate of the actual strength of anisotropy in
the ammoniated fulleride, we carried out comparative DFT-LDA
electronic structure calculations for K$_3$C$_{60}$ and a simplified 
but meaningful model of NH$_3$K$_3$C$_{60}$ using the PWSCF 
package based on plane waves and ultrasoft pseudopotentials\cite{PWSCF:note}.
We took a lattice constant $a=14.2$~\AA\ for fcc K$_3$C$_{60}$, and
$a=14.89$~\AA\ for NH$_3$K$_3$C$_{60}$, the latter with a centered
tetragonal unit cell with $c/a=0.91$, neglecting the exceedingly 
small difference between $a$ and $b$.
In both calculations a single-C$_{60}$ unit cell was used, as
pictured in Fig.~\ref{geometry_NH3:fig}. 
In these model structures and calculations, we ignored merohedral disorder,
as well as the rich antiferro-rotational structure recently discovered in
actual NH$_3$K$_3$C$_{60}$\cite{margadonna}.

We used 27~Ry and 160~Ry plane wave energy cut-offs for the 
wavefunctions and the charge density, respectively.
The Fermi surface was smeared with parameter of 0.3 eV and a
$2\times2\times2$ Monkhorst and Pack mesh \cite{Monkhorst74}.
The density of states (DOS) was calculated as a sum of Gaussians of width
$\sigma=0.03$ eV sampling $k$-space with a uniform $6\times6\times6$ mesh.

\begin{figure}[t]
\centerline{\psfig{figure=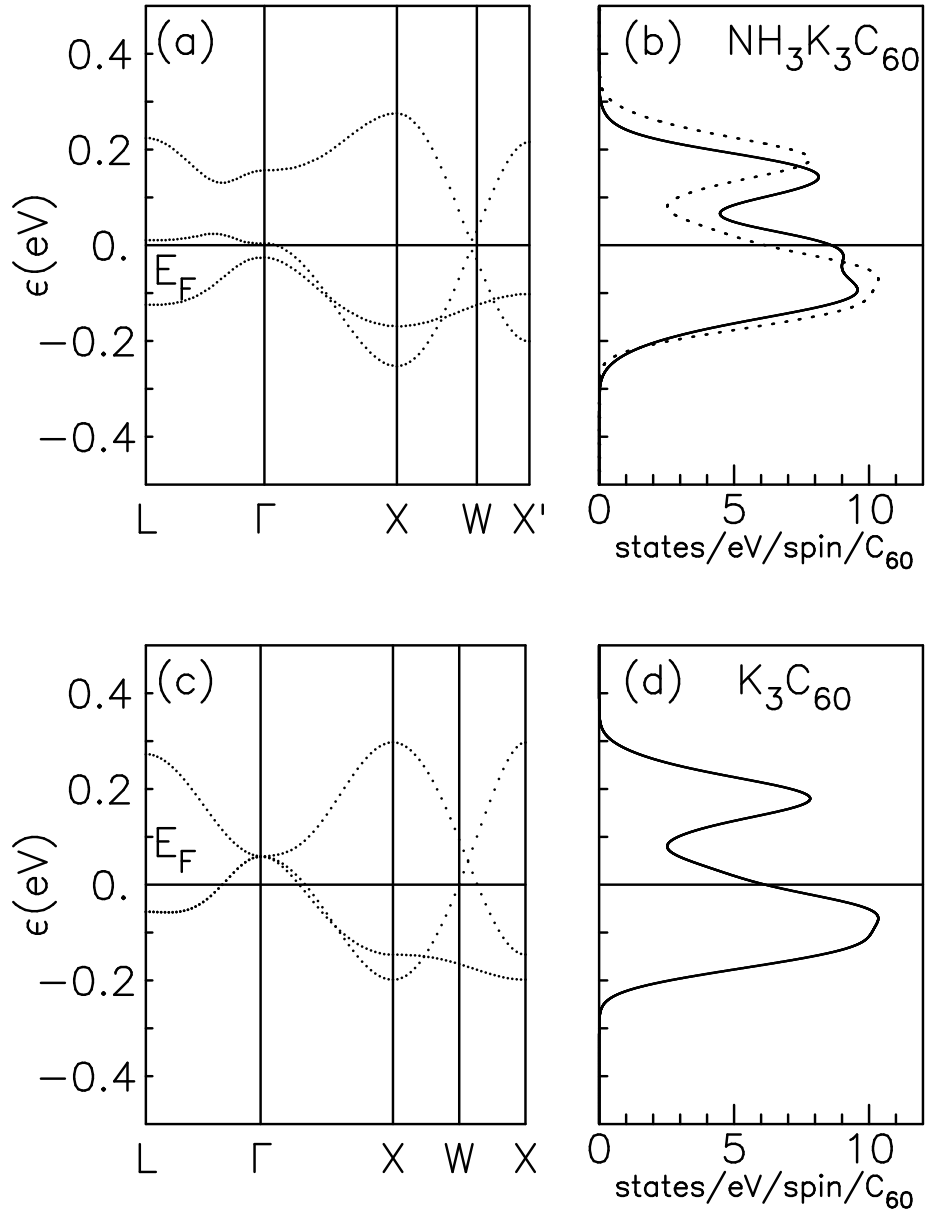,width=12.5cm}}
\vspace{0.1cm}
\caption{
Band structure of NH$_3$K$_3$C$_{60}$ (a), obtained by an DFT-LDA
calculation, compared to that of K$_3$C$_{60}$ (c).
The corresponding densities of states are given in panels (b) and (d).  
In panel (b) the density of states of K$_3$C$_{60}$ (dashed line) is re-drawn
for direct comparison with that of NH$_3$K$_3$C$_{60}$ (solid line).
\label{bands_NH3:fig}
}
\end{figure}

Fig.~\ref{bands_NH3:fig} shows the $t_{1u}$ bands of NH$_3$K$_3$C$_{60}$
(a), compared to those of K$_3$C$_{60}$ (c); the corresponding density of
states are compared in panels (b) and (d).
We find that the presence of the inserted NH$_3$ molecules modifies only
weakly the essentially pure C$_{60}$ $t_{1u}$ conduction band, as expected.
The difference in the two band structures are almost entirely due to the
different lattice structures of the molecular centers.  Small changes
appear in the details of the bands throughout the Brillouin Zone, but both
compounds are predicted in LDA to be three-band metals, their overall
$t_{1u}$ density of states not dissimilar in the ammonia-intercalated and
pristine compound.  In particular, as apparent from
Fig.~\ref{bands_NH3:fig}(b), the bandwidths of NH$_3$K$_3$C$_{60}$ and
K$_3$C$_{60}$ are close, $W\sim 0.6$~eV.
The main difference in the two band structures is a splitting of the
threefold degenerate $t_{1u}$ band of K$_3$C$_{60}$ at the $\Gamma$ point
of NH$_3$K$_3$C$_{60}$. 
This splitting is precisely a measure of the strength of the non-cubic 
crystalline environment seen by the $t_{1u}$ orbital on each fullerene
molecule in the orthorhombic structure of NH$_3$K$_3$C$_{60}$.
Its magnitude is roughly a quarter (0.2 - 0.3) of the total bandwidth;
this represents the main result of the DFT calculation. 
If this material were an uncorrelated metal, this splitting would have no
major consequences, just a change of shape of the Fermi surface. 
The consequences can be much more important due to strong correlations,
as discussed below.

\section{Discussion and conclusions} \label{discussion:sec}

The $\Gamma$-point $t_{1u}$ band splitting -- see
Fig.~\ref{bands_NH3:fig}(a) -- can be taken as a crude estimate of the
value of $\Delta \sim 0.15$~eV in NH$_3$K$_3$C$_{60}$, corresponding to a
dimensionless ratio $\Delta/W \sim 0.25$.
This value is of course too small to determine a complete 
emptying of one of the bare, uncorrelated bands, and indeed all 
three bands of NH$_3$K$_3$C$_{60}$ still cross the Fermi energy $E_{\rm F}$.
However, correlations play an important role in fullerides,
where the value of $U$ is believed to lie between 1 and 1.6~eV
\cite{Gunnarsson97}, thus substantially larger than the bandwidth $W$.
As the non-ammoniated fullerides A$_3$C$_{60}$ are still generally in the 
metallic and superconducting phase, it is not unreasonable to 
assume that their value 
of $U/W$ \cite{Antropov92,Martin93,Koch99L}
is in all likelihood only marginally smaller than the
$U_c^{\rm 3-band}$ for the Mott transition in the half-filled 
$d=3$-band cubic system. 
Indirect evidence that ordinary, uncorrelated electronic bands do not
exist, and that the electron spectral function at $E_{\rm F}$ can more
reasonably be assimilated to a narrow, dispersionless Kondo-like resonance
was provided for K$_3$C$_{60}$ by recent photoemission
data\cite{Goldoni01}.

\begin{figure}[t]
\centerline{\psfig{figure=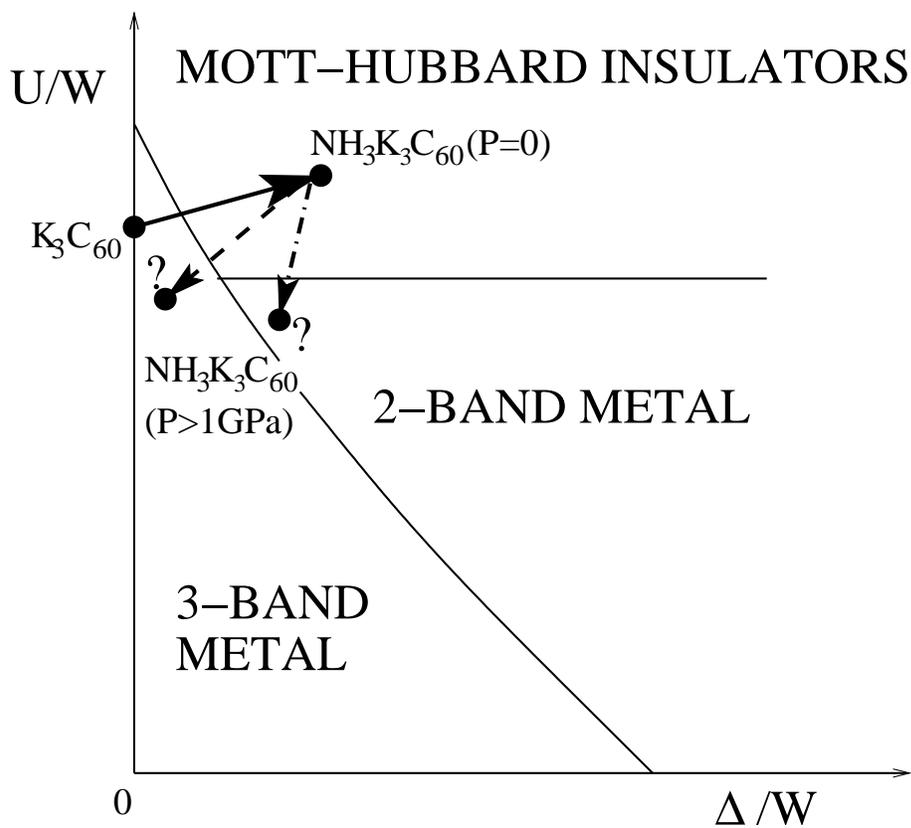,width=12.5cm}}
\vspace{0.1cm}
\caption{
A schematic $(U,\Delta)$ phase diagram for the $d=3$-bands 
family of compounds K$_3$C$_{60}$ and NH$_3$K$_3$C$_{60}$.
The dashed and dot-dashed arrows indicate two plausible paths of
pressure-induced metallization of NH$_3$K$_3$C$_{60}$.
\label{phased_NH3:fig}
}
\end{figure}

The analysis of the metal-insulator transition in our simplified two-band
model suggests a mechanism for an anisotropy-induced Mott transition in a
correlated multi-band fermionic system.
Clearly, many ingredients which we have left out of our calculation are
expected to be quantitatively important for a realistic description of the
$t_{1u}$ doped fullerides. 
However, we suggest that the mechanism which we have described in 
Secs.~\ref{model:sec} and \ref{dmft:sec}, whereby 
a small orthorhombic splitting of the bands is capable, alone, of turning a 
strongly correlated metal into a Mott insulator, could be a qualitatively 
important ingredient in the description of these systems.  
%
Specifically, we suggest that, in the appropriate $(U,\Delta)$ phase diagram 
(sketched in Fig.~\ref{phased_NH3:fig}), the
ammoniation of K$_3$C$_{60}$ into orthorhombic NH$_3$K$_3$C$_{60}$ could
correspond to a displacement\cite{Uammoniated:note} similar to that
indicated by an arrow in Fig.~\ref{phase_diagram:fig}.
Inspection of Fig.~\ref{phase_diagram:fig} shows that, if $U$ is close
enough to $U_c$, even a value of $\Delta$ substantially smaller than that
we have estimated for NH$_3$K$_3$C$_{60}$ could suffice to drive that metal
insulator transition.
%
Very recent conductivity measurements \cite{Kitano02} on the class of
compounds NH$_3$K$_{3-x}$Rb$_x$C$_{60}$ suggest indeed the orthorhombic
distortion 
as the main ingredient driving the Mott transition in these systems. 

We should stress that our picture does not exclude a role, in the
metal-insulator transition, for the accompanying ammoniation-driven
expansion of the lattice.
That role is experimentally proven by the pressure-driven insulator-metal
transition, where NH$_3$K$_3$C$_{60}$ is transformed into a metal and a
superconductor, despite the permanence of the orthorhombic
structure\cite{Margadonna01}.
However a new, interesting, question raised in Fig.~\ref{phased_NH3:fig} is
what kind of metal does one recover in NH$_3$K$_3$C$_{60}$ under pressure,
whether a three-band metal like the cubic K$_3$C$_{60}$, or a new effective
two-band metal.
The former might in effect require a smaller deformation.
It would be very interesting in this regard if the Fermi-surface topology
of NH$_3$K$_3$C$_{60}$ could be studied under pressure.
  
Another interesting -- even if perhaps not practically straightforward --
test of this overall picture could be obtained by applying uniaxial stress
to cubic superconducting fullerides of the A$_3$C$_{60}$ family.
Contrary to standard tendency of hydrostatic pressure toward metallization
and lower superconducting $T_c$, the orbital splitting associated with the
appropriate uniaxial strain could drive some of these compounds perhaps
first toward higher $T_c$, and eventually Mott insulating.
%
A similar suggestion has been recently put forward, in the context of
field-induced superconductivity \cite{schon00,Batlogg00}, by Koch
\cite{Koch02}, although on rather different grounds.
Also, recent calculations \cite{Wehrli01,Wehrli02} of the size and the effects
of the charge-inducing electric field on the 2-dimensional C$_{60}$ layer,
in the field-induced geometry, suggest that this field induces splittings
of the molecular levels which might in turn affect the electrons and
holes bands in a qualitatitevely similar way to the one we propose
here for the orthorhombic field in 3-dimensional fullerides.

In summary, we have investigated the role of a symmetry-breaking 
orbital splitting on the transition to a Mott insulating state in an 
initially orbitally degenerate strongly correlated multi-band metal.
The calculations, carried out in a highly idealized $d=2$ model, 
have unveiled the Kondo quasiparticle energy scale as that which
the non-cubic splitting magnitude must reach in order to induce the Mott
transition even at constant bandwidth and effective Coulomb repulsion.
The likely relevance of this main qualitative finding to the 
fullerides K$_3$C$_{60}$ and NH$_3$K$_3$C$_{60}$ has been highlighted. 

\section*{Acknowledgments}
We are indebted to O.\ Gunnarsson and M.\ Fabrizio, for useful discussions. 
R.\ Assaraf collaborated with this project in its earliest stage.
This work was partly supported by the European Union, contract
ERBFMRXCT970155 (TMR Fulprop), and by MIUR COFIN01, MURST COFIN99.


\begin{thebibliography}{10}

\bibitem{Brouet99}
{ V.\ Brouet, H.\ Alloul, F.\ Quere, G.\ Baumgartner, and L.\ Forro, Phys.\
  Rev.\ Lett.\ {\bf 82}, 2131 (1999)}.

\bibitem{Ramirez}
{ A.\ P.\ Ramirez, Supercond.\ Review {\bf 1}, 1 (1994)}.

\bibitem{Gunnarsson97}
{ O.\ Gunnarsson, Rev.\ Mod.\ Phys.\ {\bf 69}, 575 (1997)}.

\bibitem{schon00}
{ J.\ H.\ Sch\"on, Ch.\ Kloc, R.\ C.\ Haddon, and B.\ Batlogg, Science {\bf
  288}, 656 (2000)}.

\bibitem{Batlogg00}
{ J.\ H.\ Sch\"on, Ch.\ Kloc, and B.\ Batlogg, Nature {\bf 408}, 549 (2000)}.

\bibitem{Fabrizio97}
{ M.\ Fabrizio and E.\ Tosatti, Phys.\ Rev.\ B {\bf 55}, 13465 (1997)}.

\bibitem{Capone00}
{ M.\ Capone, M.\ Fabrizio, P.\ Giannozzi, and E.\ Tosatti, Phys.\ Rev.\ B {\bf
  62}, 7619 (2000)}.

\bibitem{Benning}
{ P.\ J.\ Benning, F.\ Stepniak, and J.\ H.\ Weaver, Phys.\ Rev.\ B {\bf 48},
  9086 (1993)}.

\bibitem{Brouet01}
{ V.\ Brouet, H.\ Alloul, T.\ N.\ Le, S.\ Garaj, L.\ Forro, Phys.\ Rev.\ Lett.\
  {\bf 86} 4680 (2001)}.

\bibitem{Rosseinsky93}
{ M.\ J.\ Rosseinsky, D.\ W.\ Murphy, R.\ M.\ Fleming, and O.\ Zhou, Nature
  {\bf 364}, 425 (1993)}.

\bibitem{Takenobu00}
{ T.\ Takenobu, T.\ Muro, Y.\ Iwasa, and T.\ Mitani, Phys.\ Rev.\ Lett.\ {\bf
  85}, 381 (2000)}.

\bibitem{Obu00}
{ T.\ T.\ Obu, H.\ Shimoda, Y.\ Iwasa, T.\ Mitani, M.\ Kosaka, K.\ U.\
  Tanigaki, C.\ M.\ Brown, and K.\ Prassides, Mol.\ Cryst.\ Liq.\ Cryst.\ {\bf
  340}, 599 (2000)}.

\bibitem{Tou1}
{ H.\ Tou, N.\ Muroga, Y.\ Maniwa, H.\ Shimoda, Y.\ Iwasa, and T.\ Mitani,
  Physica B {\bf 281}, 1018 (2000)}.

\bibitem{Zhou95}
{ O.\ Zhou, T.\ T.\ M.\ Palstra, Y.\ Iwasa, R.\ M.\ Fleming, A.\ F.\ Hebard,
  P.\ E.\ Sulewski, D.\ W.\ Murphy, and B.\ R.\ Zegarski, Phys.\ Rev.\ B {\bf
  52}, 483 (1995)}.

\bibitem{Margadonna01}
{ S.\ Margadonna, K.\ Prassides, H.\ Simoda, Y.\ Iwasa, and M.\ M\'ezouar,
  Europhys.\ Lett.\ {\bf 56}, 61 (2001)}.

\bibitem{Gunnarsson96}
{ O.\ Gunnarsson, E.\ Koch, and R.\ M.\ Martin, Phys.\ Rev.\ B {\bf 54}, R11026
  (1996)}.

\bibitem{Koch99}
{ E.\ Koch, O.\ Gunnarsson, and R.\ M.\ Martin, Phys.\ Rev.\ B {\bf 60}, 15714
  (1999)}.

\bibitem{splitting:note}
{ The origin of the band splitting at the $\Gamma$ point is twofold: (i) the
  hopping between neighbor fullerene balls is different for pairs of molecules
  separated by different distances, along different crystal directions; (ii)
  different orbitals in the molecular $t_{1u}$ manifold composing the
  conduction band acquire a different energy according to their orientation
  relative to the neighboring C$_{60}^{3-}$, NH$_3$ and K$^+$ species, forming
  the orthorhombic local environment}.

\bibitem{Georges96}
{ A.\ Georges, G.\ Kotliar, W.\ Krauth, and M.\ J.\ Rozenberg, Rev.\ Mod.\
  Phys.\ {\bf 68}, 13 (1996)}.

\bibitem{Lichtenstein01}
{ A.\ I.\ Lichtenstein, M.\ I.\ Katsnelson, and G.\ Kotliar, Phys.\ Rev.\
  Lett.\ {\bf 87}, 067205 (2001)}.

\bibitem{Capone01}
{ M.\ Capone, M.\ Fabrizio, E.\ Tosatti, Phys.\ Rev.\ Lett.\ {\bf 86}, 5361
  (2001)}.

\bibitem{MTA}
{ N.\ Manini, E.\ Tosatti, and A.\ Auerbach, Phys.\ Rev.\ B {\bf 49}, 13008
  (1994)}.

\bibitem{Lifshitz60}
{ I.\ M.\ Lifshitz, Sov.\ Phys.\ JETP {\bf 11}, 1130 (1960)}.

\bibitem{Katsnelson00}
{ M.\ I.\ Katsnelson and A.\ V.\ Trefilov, Phys.\ Rev.\ B {\bf 61}, 1643
  (2000)}.

\bibitem{Rozenberg97}
{ M.\ J.\ Rozenberg, Phys.\ Rev.\ B {\bf 55}, R4855 (1997)}.

\bibitem{Rozenberg94}
{ M.\ J.\ Rozenberg, G.\ Kotliar, and X.\ Y.\ Zhang, Phys.\ Rev.\ B {\bf 49},
  10181-10193 (1994)}.

\bibitem{Caffarel94}
{ M.\ Caffarel and W.\ Krauth, Phys.\ Rev.\ Lett.\ {\bf 72}, 1545 (1994)}.

\bibitem{Bulla99}
{ R.\ Bulla, Phys.\ Rev.\ Lett.\ {\bf 83}, 136 (1999)}.

\bibitem{Arovas95}
{ D.\ P.\ Arovas and A.\ Auerbach Phys.\ Rev.\ B {\bf 52}, 10114-10121 (1995)}.

\bibitem{Sutherland75}
{ B.\ Sutherland, Phys.\ Rev.\ B {\bf 12}, 3795 (1975)}.

\bibitem{Zhang98}
{ Y.\ Q.\ Li, Michael Ma, D.\ N.\ Shi, and F.\ C.\ Zhang, Phys.\ Rev.\ Lett.\
  {\bf 81}, 3527 (1998)}.

\bibitem{DMFTdetails:note}
{ Our implementation of the DMFT follows closely that of
  Ref.~\protect\cite{Caffarel94}. The chemical potential is adjusted at each
  iteration, so that the desired number of electrons per site is obtained at
  convergence\protect\cite{Costi02}. We find advantageous to use a logarithmic
  mesh for the imaginary frequencies, with a low-frequency cutoff of the order
  of $0.02\;W$, to insure convergence with the rather small number of baths per
  band we can include ($n_b=4\div 5$). Our results are in substantial accord
  with those of the traditional linear mesh}.

\bibitem{DMFTMCdetails:note}  	
{ Working at zero temperature is a quite important feature in our
  specific case, where the competition between the splitting $\Delta$
  and the small quasi-particle residue $z$ might be blurred,
  within a standard Monte Carlo approach, by finite temperature effects.
  The exact diagonalization method suffers, however, from a severe constraint
  on the number of discretized conduction band states which can be included.
  As a consequence, for realistic calculations with three or more degenerate 
  orbitals a Monte Carlo approach is more appropriate \cite{Georges96}}.

\bibitem{PWSCF:note}
{ The package, togheter with the ultrasoft pseudopotentials of C, N, H and
  local potential of K, can be found at the URL http://www.pwscf.org}.

\bibitem{margadonna}
{ S.\ Margadonna, K.\ Prassides, H.\ Shimoda, T.\ Takenobu, Y.\ Iwasa, Phys.\
  Rev.\ B {\bf 64} 132414 (2001)}.

\bibitem{Monkhorst74}
{ H.\ J.\ Monkhorst and J.\ D.\ Pack, Phys.\ Rev.\ B {\bf 13}, 5188 (1974)}.

\bibitem{Antropov92} 
V.\ P.\ Antropov, O.\ Gunnarsson, and O.\ Jepsen,
Phys.\ Rev.\ B {\bf 46}, 13647 (1992).

\bibitem{Martin93} 
R.\ L.\ Martin and J.\ P.\ Ritchie, Phys.\ Rev.\ B {\bf 48}, 4845 (1993).

\bibitem{Koch99L}
E.\ Koch, O.\ Gunnarsson, and R.\ M.\ Martin, Phys.\ Rev.\ Lett.\ {\bf 83}, 620
(1999).

\bibitem{Goldoni01}
{ A.\ Goldoni, L.\ Sangaletti, F.\ Parmigiani, G.\ Comelli and G.\ Paolucci,
  Phys.\ Rev.\ Lett. {\bf 87}, 076401 (2001)}.

\bibitem{Uammoniated:note}
{ Polarization screening of the on-site interaction $U$ is different in the two
  compounds, with poorer screening in NH$_3$K$_3$C$_{60}$ leading to a larger
  effective $U/W$.\ However, a crude estimate suggests that the reduced
  screening of C$_{60}$ due to the slightly larger cell volume is probably
  over-compensated by the extra screening due to ammonia.\ As a net result, we
  expect the effective $U$ to be substantially the same in the two compounds,
  as indeed seems to be the case in different fullerides}.

\bibitem{Kitano02}
{ H.\ Kitano, R.\ Matsuo, K.\ Miwa, A.\ Maeda, T.\ Takenobu, Y.\ Iwasa, and T.\
  Mitani, Phys.\ Rev.\ Lett.\ {\bf 88}, 096401 (2002)}.

\bibitem{Koch02}
{ E.\ Koch, cond-mat/0112329}.

\bibitem{Wehrli01}
{ S.\ Wehrli, D.\ Poilblanc, and T.\ M.\ Rice, Eur.\ Phys.\ J.\ B {\bf 23}, 345
  (2001)}.

\bibitem{Wehrli02}
{ S.\ Wehrli, private communication}.

\bibitem{Costi02}
{ T.\ Costi and N.\ Manini, J.\ Low Temp.\ Phys.\ {\bf 126}, 835 (2002)}.

\end{thebibliography}


\newpage

\end{document}